\documentclass[preprintnumbers,amsmath,amssymbm,prd]{revtex4}
\usepackage{epsfig}
\usepackage{graphicx}
\usepackage{amssymb}

\begin{document}
\title{Stationary bound-state massive scalar field configurations supported by
spherically symmetric compact reflecting stars}
\author{Shahar Hod}
\affiliation{The Ruppin Academic Center, Emeq Hefer 40250, Israel}
\affiliation{ }
\affiliation{The Hadassah Institute, Jerusalem 91010, Israel}
\date{\today}

\begin{abstract}
\ \ \ It has recently been demonstrated that asymptotically flat
neutral reflecting stars are characterized by an intriguing no-hair
property. In particular, it has been proved that these  {\it
horizonless} compact objects cannot support spatially regular {\it
static} matter configurations made of scalar (spin-0) fields, vector
(spin-1) fields, and tensor (spin-2) fields. In the present paper we
shall explicitly prove that spherically symmetric compact reflecting
stars can support {\it stationary} (rather than static) bound-state
massive scalar fields in their exterior spacetime regions. To this
end, we solve analytically the Klein-Gordon wave equation for a
linearized scalar field of mass $\mu$ and proper frequency $\omega$
in the curved background of a spherically symmetric compact
reflecting star of mass $M$ and radius $R_{\text{s}}$. It is proved
that the regime of existence of these stationary composed star-field
configurations is characterized by the simple inequalities
$1-2M/R_{\text{s}}<(\omega/\mu)^2<1$. Interestingly, in the regime
$M/R_{\text{s}}\ll1$ of weakly self-gravitating stars we derive a
remarkably compact {\it analytical} formula for the discrete
spectrum $\{\omega(M,R_{\text{s}},\mu)\}^{n=\infty}_{n=1}$ of
resonant oscillation frequencies which characterize the stationary
composed compact-reflecting-star-linearized-massive-scalar-field
configurations. Finally, we verify the accuracy of the analytically
derived resonance formula of the composed star-field configurations
with direct numerical computations.
\end{abstract}
\bigskip
\maketitle

\section{Introduction}

In their physically interesting and mathematically elegant no-hair
theorems, Bekenstein \cite{Bek1} and Teitelboim \cite{Teit} (see
also \cite{Bekto,Chas,Heu,Bek2,BekMay,Bek20} and references therein)
have explicitly proved that static non-linear matter configurations
which are made of spatially regular massive scalar fields cannot be
supported in the exterior spacetime regions of asymptotically flat
black holes \cite{Noteit,Hodrc,Herkr}. This intriguing physical
property of the static sector of the non-linearly coupled
Einstein-scalar field equations is often attributed to the fact that
classical black-hole spacetimes are characterized by one-way
membranes (event horizons) that irreversibly absorb matter and
radiation fields.

Interestingly, it has recently been demonstrated that the presence
of an absorbing (attractive) horizon is {\it not} a necessary
condition for the validity of the no-hair property for compact
objects in general relativity. Specifically, it has been explicitly
proved in \cite{Hodrec,Bha,Hodnww} that asymptotically flat
horizonless compact stars with reflecting (that is, repulsive rather
than attractive) boundary conditions are also characterized by an
intriguing no-hair property. In particular, the no-hair theorems
presented in \cite{Hodrec,Bha,Hodnww} have revealed the interesting
fact that spatially regular static configurations made of scalar
(spin-0) fields, vector (spin-1) fields, and tensor (spin-2) fields
cannot be supported by spherically symmetric compact stars with
reflecting \cite{Noteabb,Noterrb} boundary conditions.

It is of physical interest to test the general validity of the
no-hair property revealed in \cite{Hodrec,Bha,Hodnww} for {\it
horizonless} compact reflecting objects in general relativity. In
particular, in the present paper we raise the following physically
intriguing question: Can compact reflecting stars \cite{Noteref}
support {\it stationary} (rather than static) scalar fields in their
exterior spacetime regions?

In order to address this physically interesting question, we shall
analyze below the Klein-Gordon wave equation for a stationary
linearized scalar field of mass $\mu$ and proper frequency $\omega$
in the background of a spherically symmetric compact reflecting star
of mass $M$ and radius $R_{\text{s}}$. We shall first prove that the
stationary (that is, non-decaying in time) massive scalar field
configurations, {\it if} they exist, must be characterized by the
compact relation $1-2M/R_{\text{s}}<(\omega/\mu)^2<1$ \cite{Notewn}.
Interestingly, we shall then prove explicitly that spherically
symmetric compact reflecting stars {\it can} support {\it
stationary} bound-state configurations made of spatially regular
linearized massive scalar fields. In particular, we shall derive a
remarkably compact {\it analytical} formula [see Eq. (\ref{Eq34})
below] for the discrete resonant oscillation spectrum
$\{\omega(M,R_{\text{s}},\mu;n)\}^{n=\infty}_{n=1}$ \cite{Notenr0}
which characterizes the stationary composed
reflecting-star-massive-scalar-field configurations in the physical
regime $M/R_{\text{s}}\ll1$.

\section{Description of the system}

We consider a physical system which is composed of a spherically
symmetric compact reflecting star which is linearly coupled to a
stationary bound-state massive scalar field configuration. The
curved spacetime outside the compact star of mass $M$ is
characterized by the spherically symmetric line element
\cite{Chan,Noteunit}
\begin{equation}\label{Eq1}
ds^2=-f(r)dt^2+{1\over{f(r)}}dr^2+r^2(d\theta^2+\sin^2\theta
d\phi^2)\ \ \ \ \text{for}\ \ \ \ r>R_{\text{s}}\  ,
\end{equation}
where $R_{\text{s}}$ is the radius of the star and
\begin{equation}\label{Eq2}
f(r)=1-{{2M}\over{r}}\  .
\end{equation}

The dynamics of the linearized massive scalar field
$\Psi(t,r,\theta,\phi)$ in the curved spacetime of the compact star
is governed by the familiar Klein-Gordon wave equation
\cite{HodPirpam,Stro,HodCQG2,Hodch1,Hodch2,Notemm}
\begin{equation}\label{Eq3}
(\nabla^\nu\nabla_{\nu}-\mu^2)\Psi=0\  .
\end{equation}
It is convenient to decompose the scalar field eigenfunction in the
form \cite{Noteom}
\begin{equation}\label{Eq4}
\Psi(t,r,\theta,\phi)=\int\sum_{lm}e^{im\phi}S_{lm}(\theta)R_{lm}(r;\omega)e^{-i\omega
t}d\omega\  .
\end{equation}
Substituting the scalar eigenfunction (\ref{Eq4}) and the metric
components of the curved line element (\ref{Eq1}) into the
Klein-Gordon wave equation (\ref{Eq3}), one finds that the spatial
behavior of the radial scalar function $R_{lm}(r)$ is determined by
the ordinary differential equation
\cite{HodPirpam,Stro,HodCQG2,Hodch1,Hodch2,Heun,Abram}
\begin{equation}\label{Eq5}
{{d} \over{dr}}\Big[r^2f(r){{dR_{lm}}\over{dr}}\Big]+\Big[{{(\omega
r)^2}\over{f(r)}}-(\mu r)^2-K_l\Big]R_{lm}=0\  ,
\end{equation}
where $K_l=l(l+1)$ (with $l\geq |m|$) is the familiar angular
eigenvalue of the spatially regular angular scalar eigenfunction
$S_{lm}(\theta)$. Clearly, the radial scalar equation (\ref{Eq5}) is
invariant under the symmetry transformation $\omega\to -\omega$. We
shall henceforth assume, without loss of generality, that
\begin{equation}\label{Eq6}
\omega>0\  .
\end{equation}

We are interested in stationary (that is, non-decaying in time)
bound-state configurations of the massive scalar fields which are
supported by the central compact reflecting star. These field
configurations are characterized by the following two boundary
conditions: (1) an inner boundary condition of a vanishing scalar
eigenfunction \cite{Hodrec,Hodbb}
\begin{equation}\label{Eq7}
\Psi(r=R_{\text{s}})=0\
\end{equation}
at the reflecting surface of the spherically symmetric compact star,
and (2) the asymptotic large-$r$ boundary condition
\begin{equation}\label{Eq8}
\Psi(r\to\infty)\sim r^{\kappa-1}e^{-\sqrt{\mu^2-\omega^2} r}\ \ \
\text{with} \ \ \ \omega^2<\mu^2\
\end{equation}
of an exponentially decaying (normalizable) scalar eigenfunction at
spatial infinity [here $\kappa\equiv
{{M(2\omega^2-\mu^2)}/{\sqrt{\mu^2-\omega^2}}}$, see Eqs.
(\ref{Eq21})-(\ref{Eq22}) below].

The ordinary differential equation (\ref{Eq5}) for the radial scalar
eigenfunction, supplemented by the boundary conditions (\ref{Eq7})
and (\ref{Eq8}) at the surface of the spherically symmetric compact
reflecting star and at spatial infinity, determines the {\it
discrete} spectrum $\{\omega(M,R_{\text{s}},\mu,l;n)\}$
\cite{Notenr0} of resonant oscillation frequencies which
characterize the stationary bound-state massive scalar field
configurations in the curved spacetime of the central supporting
star. Below we shall use analytical techniques in order to determine
this characteristic star-field resonance spectrum.

\section{Lower bound on the resonant oscillation frequencies of the stationary bound-state
massive scalar field configurations}

In the present section we shall use simple analytical techniques in
order to prove that the stationary composed
compact-reflecting-star-massive-scalar-field configurations, if they
exist, are characterized by resonant oscillation frequencies which,
for given values of the physical parameters
$\{M,R_{\text{s}},\mu\}$, {\it cannot} be arbitrarily small. To this
end, it proves useful to define the new radial function
\cite{Noteomt}
\begin{equation}\label{Eq9}
\Phi=r^{-\delta}R\  ,
\end{equation}
in terms of which the scalar differential equation (\ref{Eq5}) takes
the form
\begin{equation}\label{Eq10}
r^2f(r){{d^2\Phi}\over{dr^2}}+\big[2\delta(r-2M)+2(r-M)\big]{{d\Phi}\over{dr}}+
\Big[\delta(\delta-1)f(r)+2\delta\big(1-{{M}\over{r}}\big)+{{(\omega
r)^2}\over{f(r)}}-(\mu r)^2-l(l+1)\Big]\Phi=0\  .
\end{equation}

Taking cognizance of Eqs. (\ref{Eq7}), (\ref{Eq8}), and (\ref{Eq9}),
one deduces that the radial eigenfunction $\Phi(r)$, which
characterizes the spatial behavior of the stationary bound-state
massive scalar field configurations in the curved spacetime
(\ref{Eq1}) of the spherically symmetric compact reflecting star,
must have (at least) one extremum point in the interval
\begin{equation}\label{Eq11}
r_{\text{ext}}\in (R_{\text{s}},\infty)\  .
\end{equation}
In particular, at this extremum point the scalar eigenfunction
$\Phi(r)$ is characterized by the simple relations
\begin{equation}\label{Eq12}
\{{{d\Phi}\over{dr}}=0\ \ \ \text{and}\ \ \
\Phi\cdot{{d^2\Phi}\over{dr^2}}<0\}\ \ \ \ \text{for}\ \ \ \
r=r_{\text{ext}}\  .
\end{equation}

Substituting the characteristic functional relations (\ref{Eq12})
into the scalar differential equation (\ref{Eq10}), one deduces that
the composed
compact-reflecting-star-stationary-bound-state-massive-scalar-field
configurations, if they exist, are characterized by the (rather
cumbersome) relation
\begin{equation}\label{Eq13}
\delta(\delta-1)f(r_{\text{ext}})+2\delta\Big(1-{{M}\over{r_{\text{ext}}}}\Big)+{{(\omega
r_{\text{ext}})^2}\over{f(r_{\text{ext}})}}-(\mu
r_{\text{ext}})^2-l(l+1)>0\
\end{equation}
at the extremum point $r=r_{\text{ext}}$. Taking cognizance of Eq.
(\ref{Eq2}), one finds that the inequality (\ref{Eq13}) yields the
upper bound
\begin{equation}\label{Eq14}
{{(\omega r_{\text{ext}})^2}\over{f(r_{\text{ext}})}}>(\mu
r_{\text{ext}})^2+l(l+1)-\delta^2f(r_{\text{ext}})-\delta\
\end{equation}
on the resonant oscillation frequencies of the stationary
bound-state massive scalar field configurations. The r.h.s of
(\ref{Eq14}) is maximized for the simple choice
$\delta=-[2f(r_{\text{ext}})]^{-1}$, in which case one obtains from
(\ref{Eq14}) the simple lower bound
\begin{equation}\label{Eq15}
\omega^2>\Big(1-{{2M}\over{r_{\text{ext}}}}\Big)\Big[\mu^2+{{l(l+1)}\over{r^2_{\text{ext}}}}\Big]
+{{1}\over{4r^2_{\text{ext}}}}\  .
\end{equation}

Finally, using the inequality $r_{\text{ext}}>R_{\text{s}}$ in
(\ref{Eq15}) and taking cognizance of the relation (\ref{Eq8}), one
finds that the characteristic resonant oscillation frequencies of
the stationary bound-state massive scalar field configurations in
the curved spacetime of the spherically symmetric compact reflecting
star are restricted to the dimensionless frequency interval
\cite{Notewn}
\begin{equation}\label{Eq16}
1-{{2M}\over{R_{\text{s}}}}<\Big({{\omega}\over{\mu}}\Big)^2<1\  .
\end{equation}

\section{The characteristic resonance condition of the stationary composed
compact-reflecting-star-linearized-massive-scalar-field
configurations}

Interestingly, in the present section we shall explicitly show that
the ordinary differential equation (\ref{Eq5}), which determines the
spatial behavior of the stationary bound-state massive scalar fields
in the spacetime of the compact reflecting star, is amenable to an
{\it analytical} treatment in the large-radii regime \cite{Notesw}
\begin{equation}\label{Eq17}
R_{\text{s}}\gg M\  .
\end{equation}
In particular, we shall now derive the fundamental resonance
condition which determines the {\it discrete} spectrum
$\{\omega(M,R_{\text{s}},\mu,l;n)\}$ of resonant oscillation
frequencies that characterize the composed
reflecting-star-stationary-scalar-field configurations in the
physical regime (\ref{Eq17}).

Substituting the composed radial function
\begin{equation}\label{Eq18}
\psi=rf^{1/2}(r)R\
\end{equation}
into the scalar radial equation (\ref{Eq5}) and neglecting terms of
order $O(M/r^3)$ in the large-$r$ regime (\ref{Eq17}), one obtains
the characteristic Schr\"odinger-like differential equation
\begin{equation}\label{Eq19}
{{d^2\psi}\over{dr^2}}+\Big[\omega^2-\mu^2+{{2M(2\omega^2-\mu^2)}\over{r}}+{{4M^2(3\omega^2-\mu^2)-l(l+1)}
\over{r^2}}\Big]\psi=0\
\end{equation}
which determines the spatial behavior of the massive scalar field
configurations.

It is convenient to use the dimensionless radial coordinate
\begin{equation}\label{Eq20}
x=2\sqrt{\mu^2-\omega^2}r\  ,
\end{equation}
in terms of which the scalar radial equation (\ref{Eq19}) takes the
compact form
\begin{equation}\label{Eq21}
{{d^2\psi}\over{dx^2}}+\Big(-{1\over4}+{{\kappa}\over{x}}+{{{1\over4}+\beta^2}
\over{x^2}}\Big)\psi=0\  ,
\end{equation}
where the dimensionless physical parameters $\{\kappa,\beta\}$,
which characterize the composed star-field system, are given by
\begin{equation}\label{Eq22}
\kappa\equiv {{M(2\omega^2-\mu^2)}\over{\sqrt{\mu^2-\omega^2}}}\
\end{equation}
and
\begin{equation}\label{Eq23}
\beta^2\equiv 4M^2(3\omega^2-\mu^2)-(l+{1\over2})^2\ .
\end{equation}
The general mathematical solution of the Whittaker differential
equation (\ref{Eq21}) can be expressed in terms of the familiar
confluent hypergeometric functions (see Eqs. 13.1.32 and 13.1.33 of
\cite{Abram}):
\begin{equation}\label{Eq24}
\psi(x)=e^{-{1\over2}x}x^{{1\over2}+i\beta}\big[A\cdot
U({1\over2}+i\beta-\kappa,1+2i\beta,x)+B\cdot
M({1\over2}+i\beta-\kappa,1+2i\beta,x)\big]\ ,
\end{equation}
where $\{A,B\}$ are normalization constants.

The asymptotic spatial behavior of the radial function (\ref{Eq24})
is given by (see Eqs. 13.1.4 and 13.1.8 of \cite{Abram})
\begin{equation}\label{Eq25}
\psi(x\to\infty)=A\cdot x^{\kappa}e^{-{1\over2}x}+B\cdot
{{\Gamma(1+2i\beta)}\over{\Gamma({1\over2}+i\beta-\kappa)}}x^{-\kappa}e^{{1\over2}x}\
.
\end{equation}
Taking cognizance of the boundary condition (\ref{Eq8}), which
characterizes the asymptotic spatial behavior of the stationary
bound-state (normalizable) massive scalar field configurations, one
deduces that the coefficient of the exponentially diverging term in
the asymptotic radial expression (\ref{Eq25}) should vanish:
\begin{equation}\label{Eq26}
B=0\  .
\end{equation}
One therefore finds that, in the physical regime $R_{\text{s}}\gg M$
of weakly self-gravitating stars [see (\ref{Eq17})], the supported
bound-state configurations of the stationary massive scalar fields
are characterized by the radial eigenfunction
\begin{equation}\label{Eq27}
\psi(x)=A\cdot e^{-{1\over2}x}x^{{1\over2}+i\beta}
U({1\over2}+i\beta-\kappa,1+2i\beta,x)\ \ \ \ \text{for}\ \ \ \
R_{\text{s}}\gg M\  ,
\end{equation}
where $U(a,b,z)$ is the confluent hypergeometric function of the
second kind \cite{Abram}.

Defining the dimensionless physical parameters
\begin{equation}\label{Eq28}
\alpha\equiv M\mu\ \ \ \ ; \ \ \ \ \ell\equiv l+{1\over2}\ \ \ \ ; \
\ \ \ \gamma\equiv\mu R_{\text{s}} \ \ \ \ ; \ \ \ \ \varpi\equiv
{{\omega}\over{\mu}}\  ,
\end{equation}
and taking cognizance of the inner boundary condition (\ref{Eq7})
which characterizes the behavior of the massive scalar fields at the
reflecting surface of the central compact star, one obtains from the
radial solution (\ref{Eq27}) the non-linear resonance condition
\begin{equation}\label{Eq29}
U\Big({1\over2}+i\sqrt{4\alpha^2(3\varpi^2-1)-\ell^2}-{{\alpha(2\varpi^2-1)}\over{\sqrt{1-\varpi^{2}}}},
1+2i\sqrt{4\alpha^2(3\varpi^2-1)-\ell^2},2\gamma\sqrt{1-\varpi^2}\Big)=0\
\end{equation}
for the composed
spherically-symmetric-reflecting-star-linearized-massive-scalar-field
configurations.

Interestingly, as we shall explicitly show below, the mathematically
compact resonance equation (\ref{Eq29}), which is valid in the
large-radii regime (\ref{Eq17}), determines the {\it discrete}
family of resonant oscillation frequencies
$\{\varpi(\alpha,\gamma,\ell;n)\}$ which characterize the stationary
bound-state massive scalar field configurations in the curved
spacetime of the spherically symmetric compact reflecting star.

\section{The characteristic resonance spectrum of the stationary composed
compact-reflecting-star-massive-scalar-field configurations}

The analytically derived resonance condition (\ref{Eq29}), which
determines the resonant oscillation spectrum of the stationary
bound-state linearized massive scalar fields in the background of
the compact reflecting star, can easily be solved numerically.
Interestingly, one finds that a composed
compact-reflecting-star-massive-scalar-field system with a given set
of the dimensionless physical parameters $\{\alpha,\gamma,\ell\}$ is
characterized by a discrete resonant oscillation spectrum of the
form \cite{Noteopm}
\begin{equation}\label{Eq30}
\sqrt{1-{{2M}/{R_{\text{s}}}}}<\varpi^{\text{min}}\equiv\varpi_1<\varpi_2<\varpi_3<\cdots
<\varpi_{\infty}\
\end{equation}
with the asymptotic property $\varpi_{\infty}\to 1$ [see Eq.
(\ref{Eq34}) below].

In Table \ref{Table1} we display, for various values of the
dimensionless star-field mass parameter $\alpha\equiv M\mu$, the
value of the smallest resonant oscillation frequency
$\varpi^{\text{min}}(\alpha)$. Interestingly, one finds that the
resonant oscillation frequency $\varpi^{\text{min}}(\alpha)$, which
characterizes the composed stationary
compact-reflecting-star-massive-scalar-field configurations in the
large-radii regime (\ref{Eq17}), is a monotonically decreasing
function of the dimensionless physical parameter $\alpha$. As a
consistency check, we note that the numerically computed resonant
oscillation frequencies $\varpi^{\text{min}}(\alpha)$ of the
composed star-field system conform to the analytically derived
bounds $\sqrt{1-2M/R_{\text{s}}}<\varpi<1$ [see Eq. (\ref{Eq16})].

\begin{table}[htbp]
\centering
\begin{tabular}{|c|c|c|c|c|c|c|c|}
\hline $\text{\ Dimensionless mass parameter\ \ } \alpha\equiv M\mu\
$ & \ $\ 0.1\ $\ \ & \ $\ 0.3\ $\ \ & \
$\ 0.5\ $\ \ \ & \ $\ 0.7\ $\ \ & \ $\ 0.9\ $\ \ & \ $\ 1.1\ $\ \ \\
\hline \ $\ \ \varpi^{\text{min}}(\alpha;\gamma=10,l=0)\ $\ \ \ &\ \
0.9982\ \ \ &\ \ 0.9909\ \ \ &\ \ 0.9816\ \ \ &
\ \ 0.9711\ \ \ &\ \ 0.9599\ \ \ &\ \ 0.9483\ \ \ \\
\hline
\end{tabular}
\caption{Resonant oscillation frequencies of the composed stationary
compact-reflecting-star-linearized-massive-scalar-field
configurations with $\gamma=10$ and $l=0$. We present the smallest
resonant oscillation frequency $\varpi^{\text{min}}(\alpha)$ of the
stationary bound-state massive scalar fields for various values of
the dimensionless star-field mass parameter $\alpha\equiv M\mu$. One
finds that the characteristic resonant frequency
$\varpi^{\text{min}}(\alpha)$ of the composed star-field system is a
monotonically {\it decreasing} function of the dimensionless mass
parameter $\alpha$. The resonant oscillation frequencies conform to
the analytically derived bounds $\sqrt{1-2M/R_{\text{s}}}<\varpi<1$
[see Eq. (\ref{Eq16})].} \label{Table1}
\end{table}

In Table \ref{Table2} we present, for various values of the
dimensionless mass-radius parameter $\gamma\equiv \mu R_{\text{s}}$
of the composed star-field system, the value of the smallest
resonant oscillation frequency $\varpi^{\text{min}}(\gamma)$ which
characterizes the stationary bound-state field configurations. One
finds that the characteristic resonant oscillation frequency
$\varpi^{\text{min}}(\gamma)$ of the stationary composed
compact-reflecting-star-massive-scalar-field configurations is a
monotonically increasing function of the dimensionless mass-radius
parameter $\gamma$.

\begin{table}[htbp]
\centering
\begin{tabular}{|c|c|c|c|c|c|c|c|}
\hline $\text{\ Dimensionless mass-radius\ \ } \gamma\equiv \mu
R_{\text{s}}\ $ & \ $\ 1\ $\ \ & \ $\ 2\ $\ \ & \
$\ 3\ $\ \ \ & \ $\ 4\ $\ \ & \ $\ 5\ $\ \ & \ $\ 6\ $\ \ \\
\hline \ $\ \ \varpi^{\text{min}}(\gamma;\alpha=0.1,l=0)\ $\ \ \ &\
\ 0.9959\ \ \ &\ \ 0.9966\ \ \ &\ \ 0.9970\ \ \ &
\ \ 0.9973\ \ \ &\ \ 0.9975\ \ \ &\ \ 0.9977\ \ \ \\
\hline
\end{tabular}
\caption{Resonant oscillation frequencies of the composed stationary
compact-reflecting-star-linearized-massive-scalar-field
configurations with $\alpha=0.1$ and $l=0$. We display the smallest
resonant oscillation frequency $\varpi^{\text{min}}(\gamma)$ of the
stationary bound-state massive scalar fields for various values of
the dimensionless mass-radius parameter $\gamma\equiv \mu
R_{\text{s}}$. One finds that the characteristic resonant frequency
$\varpi^{\text{min}}(\gamma)$ of the composed star-field system is a
monotonically {\it increasing} function of the dimensionless
mass-radius parameter $\gamma$. As expected, the resonant
oscillation frequencies conform to the analytically derived bounds
$\sqrt{1-2M/R_{\text{s}}}<\varpi<1$.} \label{Table2}
\end{table}

In Table \ref{Table3} we display, for various values of the
dimensionless angular harmonic index $l$ of the stationary scalar
field modes, the value of the smallest resonant oscillation
frequency $\varpi^{\text{min}}(l)$. Interestingly, one finds that
the resonant oscillation frequency $\varpi^{\text{min}}(l)$ of the
stationary bound-state massive scalar field configurations is a
monotonically increasing function of the spherical harmonic index
$l$. Again, one finds that the numerically computed resonant
oscillation frequencies $\varpi^{\text{min}}(l)$ of the composed
star-field system conform to the analytically derived bounds
$\sqrt{1-2M/R_{\text{s}}}<\varpi<1$.

\begin{table}[htbp]
\centering
\begin{tabular}{|c|c|c|c|c|c|c|c|}
\hline $\text{\ Angular harmonic index\ } l\ $ & \ $\ 0\ $\ \ & \ $\
1\ $\ \ & \
$\ 2\ $\ \ \ & \ $\ 3\ $\ \ & \ $\ 4\ $\ \ & \ $\ 5\ $\ \ \\
\hline \ $\ \ \varpi^{\text{min}}(l;\alpha=1,\gamma=10)\ $\ \ \ &\ \
0.9541\ \ \ &\ \ 0.9569\ \ \ &\ \ 0.9622\ \ \ &
\ \ 0.9694\ \ \ &\ \ 0.9773\ \ \ &\ \ 0.9841\ \ \ \\
\hline
\end{tabular}
\caption{Resonant oscillation frequencies of the composed stationary
compact-reflecting-star-linearized-massive-scalar-field
configurations with $\alpha=1$ and $\gamma=10$. We present the
smallest resonant oscillation frequency $\varpi^{\text{min}}(l)$
which characterizes the stationary bound-state star-field
configurations for various values of the dimensionless harmonic
index $l$. One finds that the characteristic resonant frequency
$\varpi^{\text{min}}(l)$ of the composed star-field system is a
monotonically {\it increasing} function of the angular harmonic
index $l$. As a consistency check, we note that the numerically
computed resonant oscillation frequencies conform to the
analytically derived bounds $\sqrt{1-2M/R_{\text{s}}}<\varpi<1$ [see
Eq. (\ref{Eq16})].} \label{Table3}
\end{table}

\section{Analytical treatment for marginally-bound stationary scalar field resonances}

In the present section we shall explicitly show that the resonance
equation (\ref{Eq29}), which determines the discrete family of
resonant oscillation frequencies $\{\varpi(\alpha,\gamma,\ell;n)\}$
that characterize the stationary bound-state linearized massive
scalar fields in the curved spacetime of the spherically symmetric
compact reflecting star, can be solved {\it analytically} in the
asymptotic regime \cite{Noteamm}
\begin{equation}\label{Eq31}
\varpi^2\to 1^{-}\  .
\end{equation}

Substituting the asymptotic relation (see Eq. 13.5.16 of
\cite{Abram})
\begin{equation}\label{Eq32}
U(a,b,x)\simeq
\Gamma({1\over2}b-a+{1\over4})\pi^{-{1\over2}}e^{{1\over2}x}x^{{1\over4}-{1\over2}b}
\cdot\cos[\sqrt{2(b-2a)x}+\pi(a-{1\over2}b+{1\over4})]\ \ \ \
\text{for}\ \ \ \ a\to -\infty\
\end{equation}
of the confluent hypergeometric function into the analytically
derived resonance condition (\ref{Eq29}), one finds the
characteristic equation
\begin{equation}\label{Eq33}
\cos\Big(\sqrt{8\alpha\gamma}-
{{\alpha\pi}\over{\sqrt{1-\varpi^2}}}+{1\over4}\pi\Big)=0\
\end{equation}
for the marginally-bound ($\varpi^2\to 1^{-}$) stationary scalar
field configurations. The resonance condition (\ref{Eq33}) yields
the compact expression \cite{Notenr}
\begin{equation}\label{Eq34}
\varpi^2=1-\Big({{\alpha}\over{n-{1\over4}+c}}\Big)^2\ \ \
\text{with}\ \ \ c\equiv{{\sqrt{8\alpha\gamma}}\over{\pi}}\ \ \ ; \
\ \ n\gg1\
\end{equation}
for the discrete spectrum of resonant oscillation frequencies which
characterize the marginally-bound composed
spherically-symmetric-compact-reflecting-star-stationary-massive-scalar-field
configurations.

It is worth emphasizing the fact that, in accord with the exact
({\it numerically} computed) data presented in Tables \ref{Table1}
and \ref{Table2}, the {\it analytically} derived expression
(\ref{Eq34}) for the asymptotic resonance spectrum implies that the
resonant oscillation frequencies which characterize the stationary
composed compact-reflecting-star-massive-scalar-field configurations
are a monotonically decreasing function of the dimensionless
star-field mass parameter $\alpha\equiv M\mu$ and a monotonically
increasing function of the dimensionless mass-radius parameter
$\gamma\equiv \mu R_{\text{s}}$.

\section{Numerical confirmation}

It is of physical interest to test the accuracy of the approximated
({\it analytically} derived) compact formula (\ref{Eq34}) for the
asymptotic resonance spectrum which characterizes the stationary
marginally-bound composed
compact-reflecting-star-massive-scalar-field configurations.

In Table \ref{Table4} we present the resonant oscillation
frequencies $\varpi^{\text{analytical}}(n;\alpha,\gamma,\ell)$ of
the stationary massive scalar fields as obtained from the
approximated (analytical) resonance formula (\ref{Eq34}) in the
asymptotic regime (\ref{Eq31}) of marginally-bound star-field
configurations. Also displayed in Table \ref{Table4} are the
corresponding resonant oscillation frequencies
$\varpi^{\text{numerical}}(n;\alpha,\gamma,\ell)$ of the stationary
bound-state massive scalar field configurations as obtained by
solving numerically the analytically derived resonance equation
(\ref{Eq29}). In the physical regime (\ref{Eq31}) of
marginally-bound star-field configurations, one finds a remarkably
good agreement \cite{Notega} between the exact resonant oscillation
frequencies [as computed {\it numerically} directly from the
resonance equation (\ref{Eq29})] and the corresponding {\it
analytically} derived resonant oscillation frequencies (\ref{Eq34})
which characterize the stationary composed
compact-reflecting-star-massive-scalar-field configurations.

\begin{table}[htbp]
\centering
\begin{tabular}{|c|c|c|c|c|c|c|c|c|}
\hline \text{Formula} & \ $\varpi(n=1)$\ \ & \ $\varpi(n=2)$\ \ & \
$\varpi(n=3)$\ \ & \ $\varpi(n=4)$\ \ & \ $\varpi(n=5)$\ \ & \
$\varpi(n=6)$\ \ & \ $\varpi(n=7)$
\ \ & \ $\varpi(n=8)$\ \ \\
\hline \ {\text{Analytical}}\ [Eq. (\ref{Eq34})]\ \ &\ 0.96058\ \ &\
0.97605\ \ &\ 0.98391\ \ &\ 0.98845\ \ &\ 0.99130
\ \ &\ 0.99321\ \ &\ 0.99456\ \ &\ 0.99554\\
\ {\text{Numerical}}\ [Eq. (\ref{Eq29})]\ \ &\ 0.95415\ \ &\
0.97312\ \ &\ 0.98217\ \ &\ 0.98730\ \ &\ 0.99051
\ \ &\ 0.99264\ \ &\ 0.99412\ \ &\ 0.99521\\
\hline
\end{tabular}
\caption{Resonant oscillation frequencies of the composed stationary
compact-reflecting-star-linearized-massive-scalar-field
configurations with $\alpha=1, \gamma=10$, and $l=0$. We display the
resonant oscillation frequencies
$\varpi^{\text{analytical}}(n;\alpha,\gamma,\ell)$ of the stationary
scalar fields as obtained from the analytically derived resonance
formula (\ref{Eq34}) in the asymptotic regime (\ref{Eq31}) of
marginally-bound star-field configurations. We also present the
corresponding numerically computed resonant oscillation frequencies
$\varpi^{\text{numerical}}(n;\alpha,\gamma,\ell)$ of the stationary
composed star-field configurations. In the physical regime
(\ref{Eq31}) of marginally-bound massive scalar field
configurations, one finds a remarkably good agreement \cite{Notega}
between the exact resonant field frequencies [as obtained {\it
numerically} from the characteristic resonance condition
(\ref{Eq29})] and the corresponding {\it analytically} derived
resonant oscillation frequencies (\ref{Eq34}) which characterize the
stationary bound-state star-field configurations.} \label{Table4}
\end{table}

\section{Summary and discussion}

The elegant no-scalar-hair theorems of \cite{Bek1,Teit} (see also
\cite{Bekto,Chas,Heu,Bek2,BekMay,Bek20} and references therein) have
revealed the interesting fact that asymptotically flat classical
black-hole spacetimes with one-way membranes (absorbing event
horizons) cannot support static configurations made of massive
scalar fields. Intriguingly, it has recently been demonstrated
explicitly that the presence of an absorbing boundary (an event
horizon) is {\it not} a necessary condition for the validity of the
no-hair property for compact gravitating objects in general
relativity. In particular, it has been proved that asymptotically
flat {\it horizonless} neutral reflecting (rather than absorbing)
stars are `bald' \cite{Hodrec} in the sense that they cannot support
{\it static} bound-state scalar (spin-0) fields, vector (spin-1)
fields, and tensor (spin-2) fields \cite{Hodrec,Bha,Hodnww}.

One naturally wonders whether the no-scalar-hair behavior observed
in \cite{Hodrec} is a generic property of compact reflecting stars?
In particular, in the present paper we have raised the following
physically intriguing question: Can horizonless compact objects with
reflecting boundary conditions (as opposed to the absorbing boundary
conditions which characterize the horizons of classical black-hole
spacetimes) support {\it stationary} (rather than static) massive
scalar field configurations?

In order to address this physically interesting question, in the
present paper we have studied analytically the characteristic
Klein-Gordon wave equation for stationary linearized massive scalar
field configurations in the curved spacetime of a spherically
symmetric compact reflecting star. The main results derived in this
paper and their physical implications are as follows:

(1) It has been explicitly proved that the stationary composed
compact-reflecting-star-linearized-massive-scalar-field
configurations, if they exist, are restricted to the dimensionless
frequency regime [see Eq. (\ref{Eq16})] \cite{Noteopm}
\begin{equation}\label{Eq35}
\sqrt{1-{{2M}\over{R_{\text{s}}}}}<{{\omega}\over{\mu}}<1\  .
\end{equation}
It is worth noting that the presence of a positive lower bound on
resonant oscillation frequencies of the composed star-field
configurations agrees with the no-scalar-hair theorem presented in
\cite{Hodrec}, according to which spherically symmetric compact
reflecting stars cannot support spatially regular {\it static}
($\omega=0$) scalar field configurations.

(2) We have then proved that spherically symmetric compact
reflecting stars {\it can} support {\it stationary} (rather than
static) bound-state linearized massive scalar fields. In particular,
it has been shown that the resonant oscillation modes of the
composed compact-reflecting-star-massive-scalar-field system depend
on three dimensionless physical parameters: the star-field mass
parameter $\alpha\equiv M\mu$, the mass-radius parameter
$\gamma\equiv \mu R_{\text{s}}$, and the spherical harmonic index
$l$ of the field mode. Solving analytically the Klein-Gordon wave
equation for the stationary massive scalar field modes in the
large-radii regime $M/R_{\text{s}}\ll1$, we have explicitly proved
that the resonance equation [see Eqs. (\ref{Eq28}) and (\ref{Eq29})]
\begin{equation}\label{Eq36}
U\Big({1\over2}+i\sqrt{4M^2(3\omega^2-\mu^2)-(l+1/2)^2}-{{M(2\omega^2-\mu^2)}\over{\sqrt{\mu^2-\omega^2}}},
1+2i\sqrt{4M^2(3\omega^2-\mu^2)-(l+1/2)^2},2\sqrt{\mu^2-\omega^2}R_{\text{s}}\Big)=0\
\end{equation}
determines the discrete spectrum
$\{\varpi(\alpha,\gamma,\ell;n)\}_{n=1}^{n=\infty}$ of dimensionless
resonant oscillation frequencies which characterize the stationary
bound-state scalar field modes in the spacetime of the spherically
symmetric compact reflecting star.

(3) Solving numerically the analytically derived resonance condition
(\ref{Eq36}) for the composed star-field configurations, we have
demonstrated that the smallest possible resonant oscillation
frequency $\varpi^{\text{min}}$ [see Eq. (\ref{Eq30})] which
characterizes the stationary bound-state scalar field configurations
is a monotonically decreasing function of the dimensionless
star-field mass parameter $M\mu$ and a monotonically increasing
function of the dimensionless mass-radius parameter $\mu
R_{\text{s}}$ and the dimensionless angular harmonic index $l$ (see
Tables \ref{Table1}, \ref{Table2}, and \ref{Table3}).

(4) It has been explicitly proved that the large-radii resonance
equation (\ref{Eq36}) for the composed star-field configurations is
amenable to an {\it analytical} treatment in the physically
interesting regime $(\omega/\mu)^2\to 1^-$ of marginally-bound
stationary massive scalar field configurations. In particular, we
have derived the remarkably simple dimensionless analytical formula
[see Eqs. (\ref{Eq28}) and (\ref{Eq34})]
\begin{equation}\label{Eq37}
{{\omega}\over{\mu}}=\sqrt{1-\Big({{M\mu}\over{n-{1\over4}+M\mu\sqrt{8R_{\text{s}}/M\pi^2}}}\Big)^2}\
\ \ ; \ \ \ n\gg1
\end{equation}
for the discrete resonance spectrum which characterizes the
stationary marginally-bound linearized massive scalar field
configurations in the spacetime of the spherically symmetric
horizonless reflecting star.

(5) It has been explicitly demonstrated that the characteristic
resonant frequencies of the stationary marginally-bound composed
compact-reflecting-star-linearized-massive-scalar-field
configurations, as deduced from the {\it analytically} derived
resonance formula (\ref{Eq37}), agree remarkably well \cite{Notega}
with the corresponding exact resonant oscillation frequencies of the
stationary massive scalar fields as obtained {\it numerically} from
the resonance condition (\ref{Eq36}) (see Table \ref{Table4}).

Finally, we would like to emphasize again that in the present
exploration of the physical and mathematical properties of the
composed star-field system we have made two simplifying assumptions:
(1) The scalar field was treated at the linear level, and (2) the
central supporting star was treated as a weakly self-gravitating
object \cite{Notenh}. As we have explicitly proved in this paper,
using these two simplifying assumptions, one can study {\it
analytically} the physical and mathematical properties of the
stationary composed
compact-reflecting-star-linearized-massive-scalar-field system. As a
next step, it would be interesting to use {\it numerical} techniques
in order to extend our results to the physically important regime of
non-linear stationary bound-state scalar field configurations
supported by strongly self-gravitating horizonless compact stars.

\bigskip
\noindent
{\bf ACKNOWLEDGMENTS}
\bigskip

This research is supported by the Carmel Science Foundation. I would
like to thank Yael Oren, Arbel M. Ongo, Ayelet B. Lata, and Alona B.
Tea for helpful discussions.



\begin{thebibliography}{99}
\bibitem{Noterrb} It is important to note that the original no-hair theorem
presented in \cite{Hodrec} is valid for compact reflecting stars
with Dirichlet boundary conditions [that is,
$\psi(r=R_{\text{s}})=0$ at the surface $r=R_{\text{s}}$ of the
compact object]. This theorem relies on the characteristic
asymptotic behavior $\psi(r\to\infty)\to0$ of the scalar
eigenfunction for finite-mass field configurations [see Eqs. (7) and
(8) of \cite{Hodrec}] which, together with the Dirichlet boundary
condition, imply that the spatially regular static scalar fields
must have an extremum point, $r=r_{\text{peak}}\geq R_{\text{s}}$,
outside the compact star. In particular, as explicitly proved in
\cite{Hodrec}, the scalar Klein-Gordon equation is violated at this
extremum point, a fact which rules out the existence of spatially
regular static scalar hair outside the compact reflecting stars. The
no-hair theorem for the compact reflecting objects can easily be
extended to the case of Neumann boundary conditions with
$d\psi/dr=0$ at the surface $r=R_{\text{s}}$ of the star, in which
case the scalar Klein-Gordon differential equation is violated at
the surface itself. Interestingly, the regime of validity of the
no-hair property for spherically symmetric asymptotically flat
compact reflecting objects can further be extended to the more
generic case of Robin boundary conditions with $\psi+b\cdot
d\psi/dr=0$ and $b<0$ at the surface $r=R_{\text{s}}$ of the star.
In this more generic case, the radial scalar eigenfunction is
characterized by the functional relation $\psi\cdot d\psi/dr>0$ at
$r=R_{\text{s}}$, a property which, together with the characteristic
asymptotic behavior $\psi(r\to\infty)\to0$ of finite mass matter
configurations [see Eqs. (7) and (8) of \cite{Hodrec}], imply that
the scalar eigenfunction must have an extremum point,
$r=r_{\text{peak}}\geq R_{\text{s}}$, outside the reflecting star.
As explicitly proved in \cite{Hodrec}, the Klein-Gordon equation for
the static scalar field is violated at this extremum point, a fact
which rules out the existence of static spatially regular scalar
hairy configurations outside the compact reflecting objects.
\bibitem{Noteref} The term `reflecting star' is used here to describe
an object with a well defined compact surface on which the scalar
field vanishes [see Eq. (\ref{Eq7}) below].
\bibitem{Notenr0} Here the resonance parameter $n$ is an integer which characterizes
the stationary bound-state massive scalar field configurations in
the spacetime of the spherically symmetric compact reflecting star
[see Eq. (\ref{Eq30}) below].
\bibitem{Noteunit} We shall use natural units in which $G=c=\hbar=1$.
\bibitem{Notemm} Note that the physical parameter $\mu$ of the massive scalar field stands
for $\mu/\hbar$. It therefore has the dimensions of
$($length$)^{-1}$.
\bibitem{Noteom} Here the integers $l$ and $m$, which are characterized
by the relation $l\geq |m|$, are the spherical and azimuthal
harmonic indices of the stationary scalar field modes, respectively.
\bibitem{Noteomt} For brevity, we shall henceforth omit the harmonic indices $\{l,m\}$
of the stationary linearized scalar field modes.
\bibitem{Notesw} Taking cognizance of the analytically derived
inequalities $\sqrt{1-2M/R_{\text{s}}}<\omega/\mu<1$ [see Eq.
(\ref{Eq16})], one realizes that the assumed large-radii regime
(\ref{Eq17}) corresponds to stationary bound-state massive scalar
field configurations with $\omega\lesssim \mu$.
\bibitem{Noteopm} As noted above, the radial scalar equation (\ref{Eq5}) is invariant
under the symmetry transformation $\omega\to -\omega$. We shall
assume that $\omega>0$ without loss of generality.
\bibitem{Notenr} Here the integer $n\gg1$ is the resonance parameter
which characterizes the stationary composed star-field
configurations.
\bibitem{Notega} It is interesting to note that the agreement between
the analytically derived resonant oscillation frequencies of the
stationary bound-state composed star-field configurations and the
corresponding exact (numerically computed) resonant oscillation
frequencies is quite good even for the fundamental (smallest)
resonant modes with $n=O(1)$. This reasonably good
analytical/numerical agreement in the $n=O(1)$ regime is quite
surprising since the mathematically compact analytical formula
(\ref{Eq34}) for the resonant frequencies of the stationary composed
compact-reflecting-star-linearized-massive-scalar-field
configurations is formally valid for marginally-bound scalar modes
with $n\gg1$.
\end{thebibliography}
\end{document}